\RequirePackage{fix-cm}
\documentclass[smallextended]{svjour3}       
\smartqed  
\usepackage{graphicx}
\usepackage{tikz}
\usepackage[multi-part-units=single]{siunitx}
\usepackage{multirow}
\usepackage{upgreek}
\usepackage{fixltx2e}
\usepackage[misc]{ifsym}
\usepackage{float}
\usepackage[font=small,labelfont=bf,tableposition=top]{caption}
\usepackage{mathtools}
\usepackage{xcolor}
\usepackage{amsmath}
\usepackage{amssymb}

\begin{document}

\title{ 3-Dimensional Deep Learning with Spatial Erasing  for Unsupervised Anomaly Segmentation in Brain MRI
}

\titlerunning{3-Dimensional Deep Learning Methods for Unsupervised Anomaly Segmentation}        

\author{Marcel Bengs$^{1}$$^{*}$ \and Finn Behrendt$^{1}$$^{*}$ \and Julia Kr\"uger$^{2}$ \and Roland Opfer$^{2}$ \and Alexander Schlaefer$^{1}$}

\authorrunning{Bengs et al.}

\institute{\Letter \quad Marcel Bengs, \email{marcel.bengs@tuhh.de}, Tel.: +49 (0)40 42878 3389 \\ $^{*}$ Authors contributed equally  \\ $^1$ Institute of Medical Technology and Intelligent Systems, Hamburg University of Technology, Hamburg, Germany \\
$^{2}$ jung diagnostics GmbH, Germany}


\date{Preprint. Accepted for publication in IJCARS.}
\maketitle

\begin{abstract}

\textit{Purpose} 
Brain Magnetic Resonance Images (MRIs) are essential for the diagnosis of  neurological diseases. Recently, deep learning methods for unsupervised anomaly detection (UAD) have been proposed for the analysis of brain MRI. These methods rely on healthy brain MRIs and eliminate the requirement of pixel-wise annotated data compared to supervised deep learning. While a wide range of methods for UAD have been proposed, these methods are mostly 2D and only learn from MRI slices, disregarding that brain lesions are inherently 3D and the spatial context of MRI volumes remains unexploited. 

\textit{Methods} 
We investigate whether using increased spatial context by using MRI volumes combined with spatial erasing leads to improved unsupervised anomaly segmentation performance compared to learning from slices. We evaluate and compare 2D variational autoencoder (VAE) to their 3D counterpart, propose 3D input erasing, and systemically study the impact of the data set size on the performance. 

\textit{Results} Using two publicly available segmentation data sets for evaluation, 3D VAE outperform their 2D counterpart, highlighting the advantage of volumetric context. Also, our 3D erasing methods allow for further performance improvements. Our best performing 3D VAE with input erasing leads to an average DICE score of 31.40\% compared to 25.76\% for the 2D VAE. 

\textit{Conclusions} We propose 3D deep learning methods for UAD in brain MRI combined with 3D erasing and demonstrate that 3D methods clearly outperform their 2D counterpart for anomaly segmentation. Also, our spatial erasing method allows for further performance improvements and reduces the requirement for large data sets.


\end{abstract}

\section{Introduction}
\label{intro}
\begin{figure}
\centering
\includegraphics[width=1.0\textwidth]{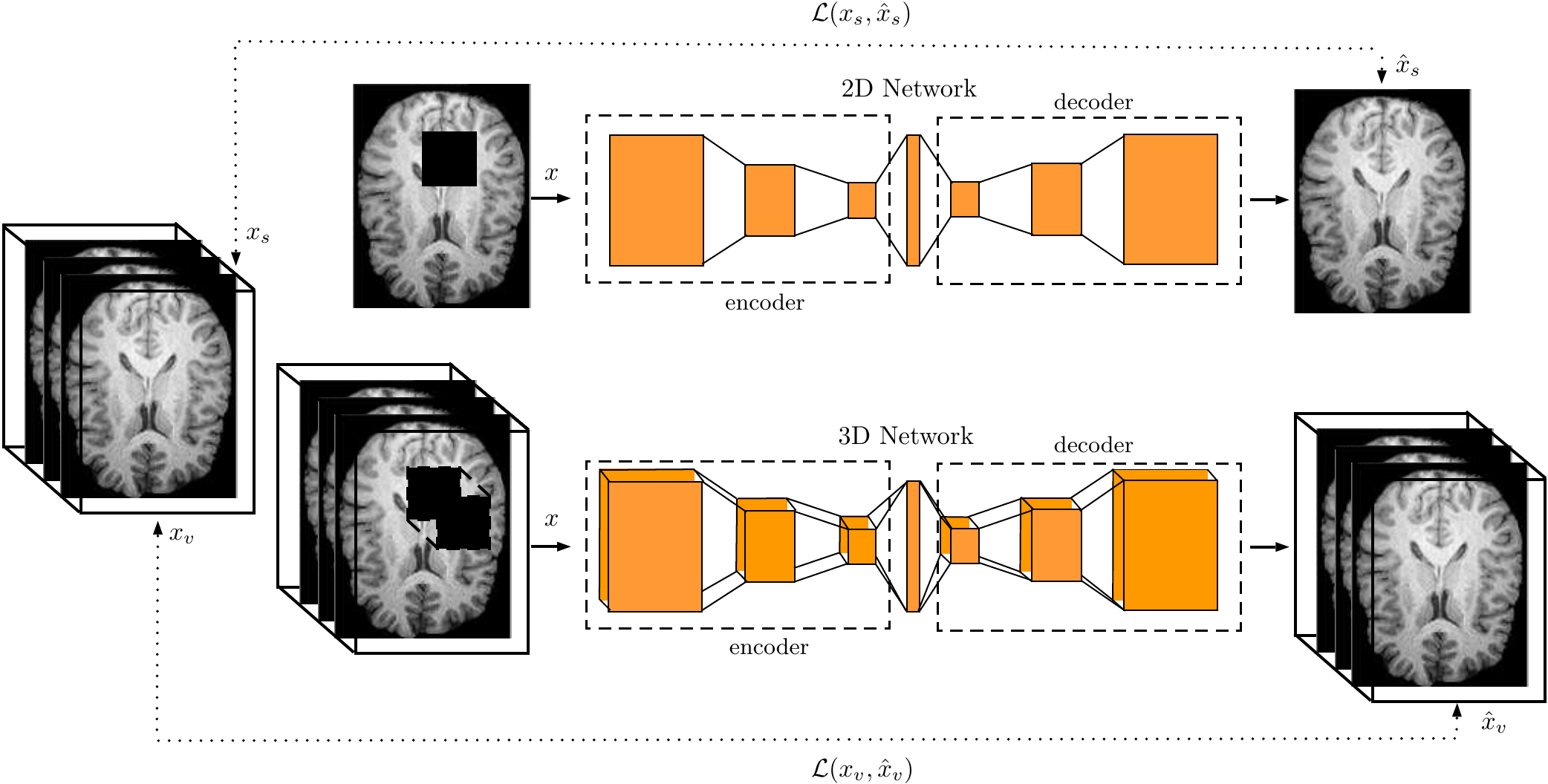}
\caption{Our approach for unsupervised anomaly segmentation using 3D deep learning combined with spatial input erasing. For the 2D network  only a single 2D slice $x_{s}$ is used as input $x$ and volumetric spatial context remains unexploited. Instead, our novel 3D approach receives an entire volume $x_{v}$ as input $x$ and learns combined features from all spatial dimensions. Also, we propose 3D spatial input erasing, where parts of the input are missing and the network is trained to restore missing image parts. Note, $\hat{x_{s}}$ and $\hat{x_{v}}$ refer to the network's reconstruction in 2D and 3D, respectively.   }
\label{fig:Motivation_Image}
\end{figure}

Brain Magnetic Resonance Images (MRIs) allow for three-dimensional (3D) imaging of the brain and are widely used in research and clinical practice for the diagnosis and treatment of neurological diseases. While promising technology advancements of the imaging quality enable an ever-increasing amount of conditions that become detectable \cite{vernooij2007incidental}, reading and interpreting MRI remains a challenging task. First, brain lesion detection and delineation requires expert knowledge and is a tedious time-consuming process, affected by human errors \cite{bruno2015understanding}. Second, MRI is increasingly used and hence an ever-increasing amount of images need to be studied, while only a limited number of experts are available \cite{chen2018unsupervised}. This leads to the urgent need for automatic detection and segmentation of lesions to assist radiologists during clinical practice.
\\
Recently, supervised deep learning methods have shown promising results for this task, while the success of these methods depends heavily on large data sets with high-quality annotations \cite{lundervold2019overview}. Note that supervised methods only generalize well to cases that are sufficiently represented in the training data. However, diverse and large annotated data sets are costly to obtain, and often only a few limited cases are available for rare diseases \cite{baur2020autoencoders}. 
\\
In contrast to that, human experts can be trained with few healthy cases to generalize, and afterwards they are able to detect even arbitrary anomalies without being trained to an explicit appearance \cite{chen2018unsupervised}. Deep learning for unsupervised anomaly detection (UAD) follows this concept of identifying unexpected, abnormal data. These methods do not require pixel-level annotations and are only trained with MRI-scans of healthy brains. Here, the task is considered as an anomaly detection problem, where the networks are trained to represent the distribution of healthy anatomy of the human brain and anomalies can be detected as outliers from the learned distribution. Typically, deep learning for UAD follows an encoder-decoder structure trained only on healthy images. Afterwards, detection and delineation of pathologies of a test image can be obtained, e.g., by pixel-wise discrepancies between the model’s input and reconstruction. \\ So far, a wide range of deep learning methods have been proposed for UAD in brain MRI, ranging from simple auto-encoders \cite{baur2018deep} to generative adversarial networks (GANs) \cite{schlegl2019f} focusing on 2D spatial information. These 2D methods have shown promising results, however, the global spatial context provided by MRI volumes remains unused and the inherently 3D structure of brains cannot be learned by the networks. This brings up the question, whether increased spatial context by using entire  MRI  volumes allows for improved performance, leading to the problem of 3D deep learning for UAD in brain MRI.  So far, 3D deep learning for UAD has hardly been considered, only pioneering work in volumetric head CT data has been proposed recently without direct comparison to 2D \cite{sato2018primitive}. 3D deep learning is challenging in nature as it results in an increased representational power that may come with an increased risk of overfitting, leading to poor generalization. For preventing the risk of overfitting several different regularization strategies have been proposed for deep learning in the context of computer vision. These methods range from simple image transformation such as rotation and flipping to adding noise during the training process, e.g., by stochastically dropping out neuron activations  \cite{srivastava2014dropout} or dropping out entire input regions \cite{devries2017improved} during training. 
Especially the latter has been combined with 2D auto-encoder networks, called context-encoders \cite{pathak2016context}, where the networks are enforced to generate the contents of an arbitrary image region conditioned on its surroundings, leading to a better understanding of the global content of the image. This idea has also shown promising results in the context of UAD in brain MRI using 2D methods \cite{zimmerer2018context} and might be a promising approach for enforcing the understanding of the global context when entire MRI volumes are used in combination with 3D deep learning. \\ \\
In this paper, we propose to learn from entire 3D MRI volumes instead of single 2D MRI slices using 3D instead of 2D unsupervised deep learning, shown in Figure \ref{fig:Motivation_Image}. Also, we extend the concept of spatial input erasing for regularization. To this end, we provide an extensive comparison of variational autoencoders (VAE) with 3D and 2D convolutions and propose several different 3D spatial erasing strategies during training. For our experiments, we use a training data set with brain MRI scans of 2008 healthy patients and evaluate our methods on two publicly available brain segmentation data sets. 
We focus on T1-weighted MRI data, which is widely used in clinics \cite{ito2019comparison,akkus2017deep}, providing a good starting point for anomaly detection. Moreover, we provide an analysis of the impact and the importance of the training data set size, especially in combination with our 3D approach.

\section{Materials and Methods}
\label{Methods}

\subsection{Data Set}
For training, we consider a data set with anonymized T1-weighted MRI volumes of 2008 healthy subjects from 22 scanners from different vendors. The resolutions in axial direction vary from \SI{0.39}{mm} to \SI{1.25}{mm} with a majority of 1310 samples with \SI{1}{mm}. The slice thickness lies between \SI{0.90}{mm} to \SI{2.40}{mm} with a majority of 906 samples with \SI{1}{mm}. 1506 samples, are acquired with a field strength of 1.5 T, 446 samples are acquired with 3 T and 56 with 1 T.
Data on all scanners was acquired during clinical routine with a standard 3D gradient echo sequence. All scans were sent to jung diagnostics GmbH for image analysis. 
\\
For evaluation, we use two publicly available data sets. First, we consider the publicly available Multimodal Brain Tumor Segmentation Challenge 2019 (BraTS 2019) data set \cite{menze2014multimodal,bakas2017advancing,bakas2018identifying} with T1-weighted image volumes of 335 subjects with the corresponding ground truth segmentation of the tumor.  The slice thickness of the BraTS 2019 data set varies from \SI{1}{mm} up to \SI{5}{mm}. Second, we use the Anatomical Tracings of Lesions After Stroke (ATLAS) data set \cite{liew2018large}, which provides T1-weighted image volumes of 304 subjects with corresponding ground truth segmentations of stroke regions. The slice thickness of the ATLAS data set varies from \SI{1}{mm} up to \SI{3}{mm}. \\
For all image volumes, we apply the following preprocessing. First, we resample all scans to the same isotropic resolution of $1\mathrm{\,mm}\times 1\mathrm{\,mm} \times 1\mathrm{\,mm}$ using cubic interpolation. Then, we follow the preprocessing of previous studies with 2D deep learning methods for UAD, which include skull stripping, denoising, and standardization \cite{baur2020autoencoders}. Next, we crop excessive background by using brain masks of the MRI scans and zero-pad all MRI scans to the largest volume resolution in our data set of $191\times158\times 163$. Last, we downsample all volumes to a size of $64\times64\times 64$ for numerical efficiency, as we encounter the computational complexity of 3D deep learning. 
Regarding our data split for training, we consider 1807 healthy images for training and 201 images for validation of our reconstruction performance. We split our data randomly and stratified by scanners. 
Considering the images of the BraTS 2019 data set, we randomly sample 133 images for validation and 202 for testing. Using the ATLAS data set, we randomly sample 121 and 183 images for validation and testing, respectively.

\subsection{Deep Learning Methods}
\label{Deep}
\begin{figure}
\centering
\includegraphics[width=1.0\textwidth]{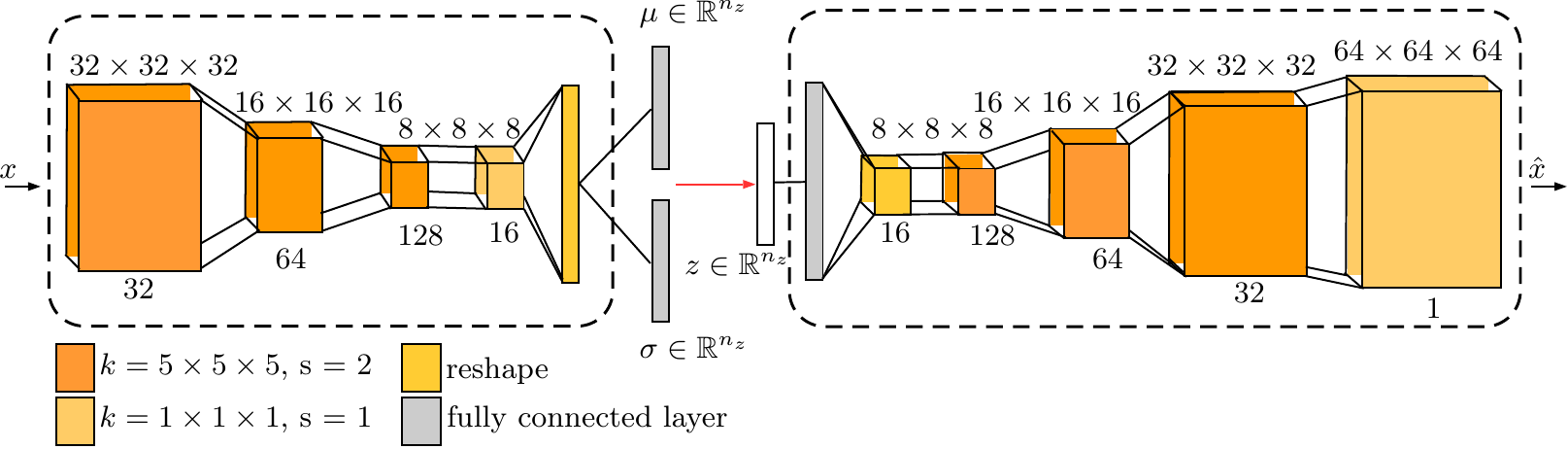}
\caption{Our backbone 3D VAD architecture receives input volume $x\in\mathbb{R}^{64\times 64 \times 64}$ and encodes it to the lower-dimensional latent variable $z\in\mathbb{R}^{n_{z}}$, afterwards the decoder reconstructs the 
 output $\hat{x}\in\mathbb{R}^{64\times 64 \times 64}$. The number over the boxes refer to the spatial size; the number below the boxes refer to the number of feature maps. We use convolutions and transposed convolutions in the encoder and decoder, respectively
 Note, the first convolution in the encoder downsamples the input from $64\times 64\times64$ to $32\times 32\times32$.} 
\label{fig:backbone_network}
\end{figure}

\begin{figure}
\centering

\includegraphics[width=0.18\textwidth]{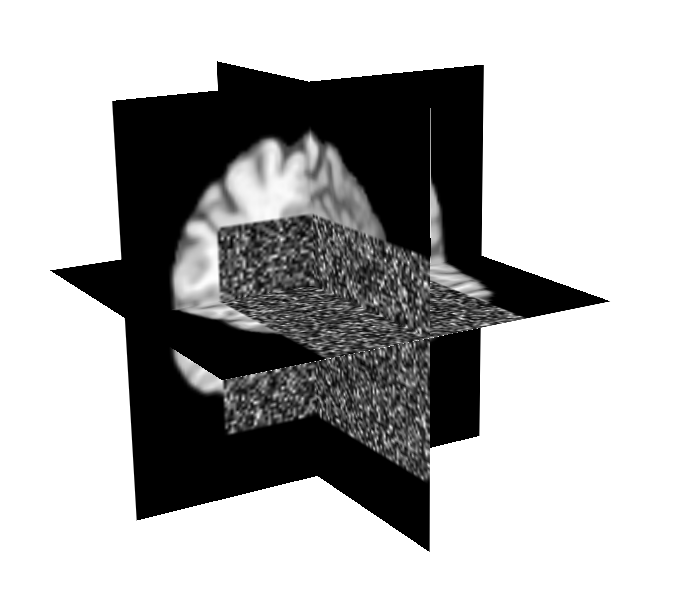}
\includegraphics[width=0.15\textwidth]{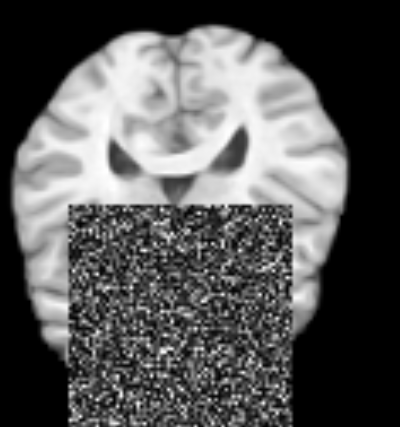}
\includegraphics[width=0.15\textwidth]{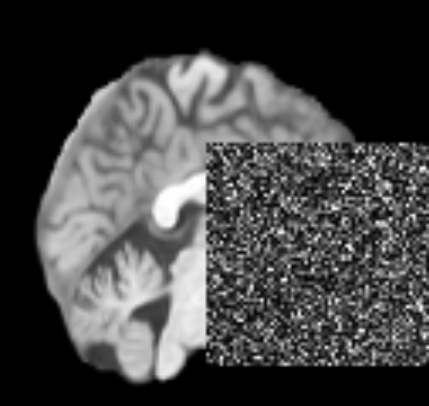}
\includegraphics[width=0.15\textwidth]{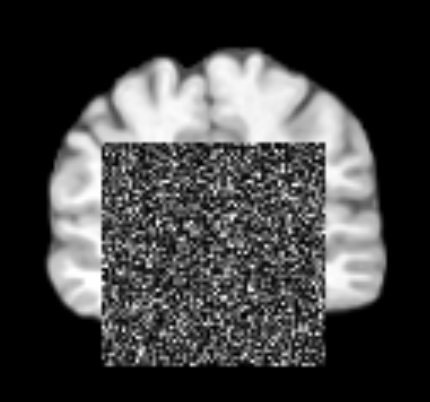}
\\
\vspace{0.5cm}
\includegraphics[width=0.18\textwidth]{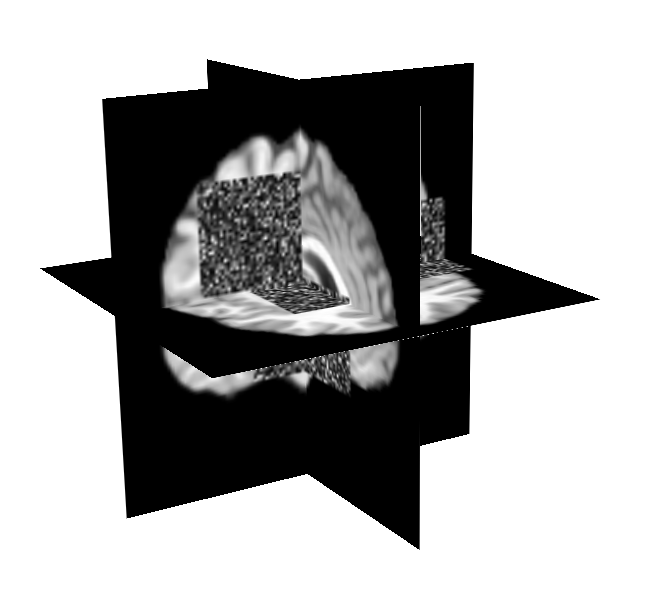}
\includegraphics[width=0.15\textwidth]{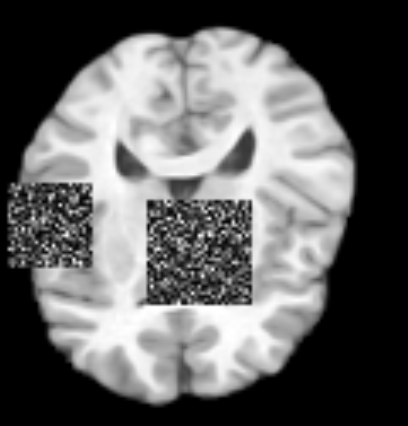}
\includegraphics[width=0.15\textwidth]{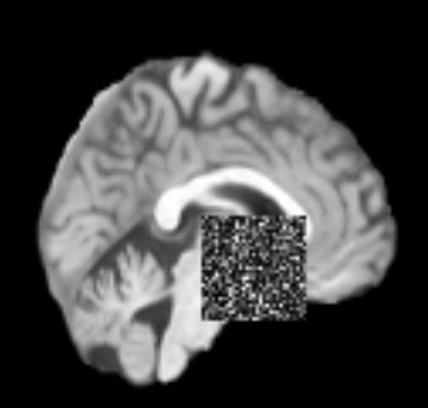}
\includegraphics[width=0.15\textwidth]{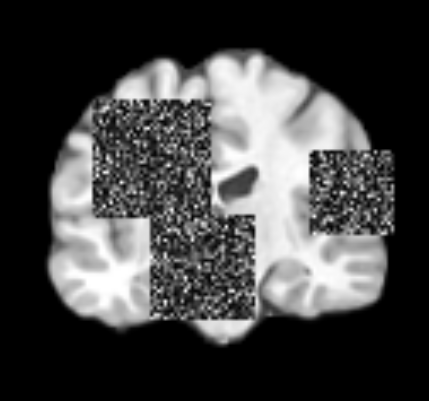} 
\\
\vspace{0.5cm}
\includegraphics[width=0.18\textwidth]{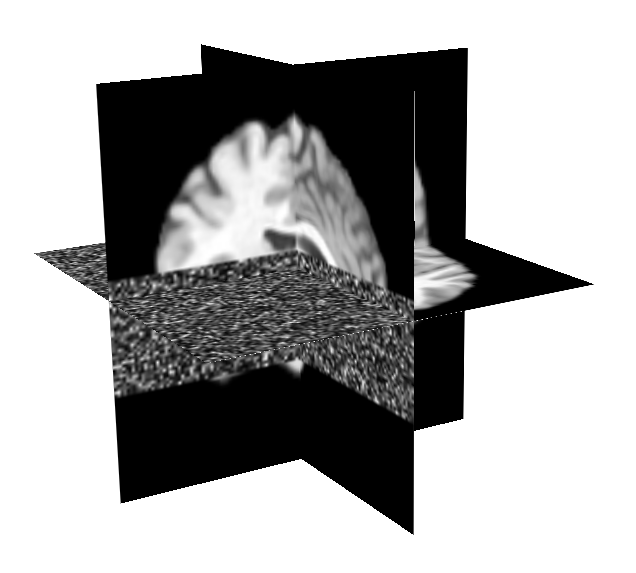}
\includegraphics[width=0.15\textwidth]{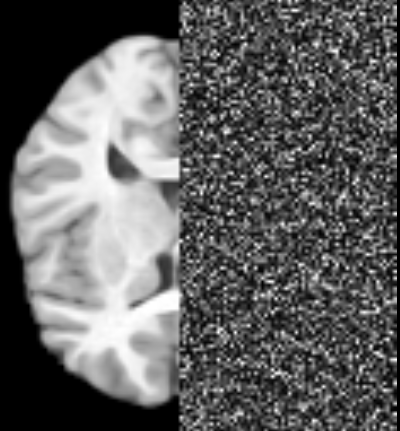}
\includegraphics[width=0.15\textwidth]{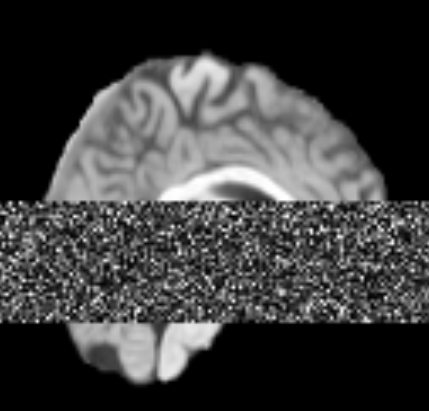}
\includegraphics[width=0.15\textwidth]{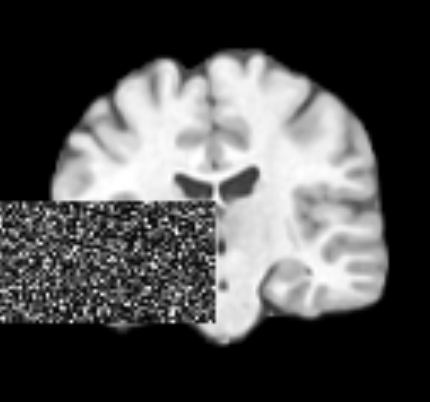}
\\

\caption{Our 3D spatial input erasing methods. In each row sectional planes of a volume with erasing are shown. Top row: We erase a single 3D cube with random location and size (Cube). Middle row: We erase multiple 3D cubes with random location and size (Multi-Cube). Bottom row: We erase an entire brain side in a subvolume (Half-Volume).}
\label{fig:earsing_methods}
\end{figure}

We address the problem of anomaly segmentation with 2D and 3D unsupervised deep learning methods using 2D MRI slices or 3D MRI volumes, respectively. Given a set of healthy MRI scans, we utilize an encoder-decoder architecture and train our methods to encode to and reconstruct from a lower-dimensional latent space $z\in\mathbb{R}^{n}$. After the methods are trained, anomalies in a test image can be detected by large reconstruction errors between the input and output image, as the networks are trained to reconstruct only images of healthy brain anatomies, e.g., fail to reconstruct abnormal image areas. \\
Recently, a comparative study on UAD using 2D deep learning methods \cite{baur2020autoencoders} has demonstrated that VAE \cite{kingma2013auto,baur2018deep} allow for promising results, while also being easy to optimize and involving fewer hyperparameters compared to other UAD methods such as GANs. Comparing the VAE with the standard AE, the VAE enforces a structure on the manifold. It has been demonstrated that this leads to performance improvements compared to the standard AE \cite{baur2020autoencoders}. Hence, we consider the concept of VAEs for our study. \\ \\
Our general backbone network is shown in Figure \ref{fig:backbone_network} and for the adaption to 2D MRI slices or 3D MRI volumes, we employ 2D or 3D operations for the network, e.g., we use 2D or 3D convolutions.  In this way, the architecture details remain the same for 2D and 3D, e.g., the number of layers and feature maps remain same, and only the dimension of the networks operation are changed. Based on our validation set performance, we choose a latent space size of $z\in\mathbb{R}^{128}$ and $z\in\mathbb{R}^{512}$ for our 2D and 3D VAE, respectively. 
\\ 
We study and extend the concept of cutout \cite{devries2017improved} and context-autoencoders \cite{pathak2016context}, which were proposed for 2D images. The main motivation behind our approach is to further enhance the usage of global image context, especially in combination with 3D methods. Therefore, we propose and evaluate the following different erasing methods for 2D and 3D, which are shown in Figure \ref{fig:earsing_methods}. Note, we only erase the regions in the input image and not in the ground-truth image that is used for optimization, hence our networks are enforced to solve an in-painting task for abnormal regions. 
\\ \\
First, we simply mask-out a single patch in the input, similar to previous concepts for 2D problems \cite{devries2017improved,pathak2016context,zimmerer2018context}. Also, we extend this approach to 3D and mask-out a single 3D cube. For the patch and cube erasing method, we randomly select a pixel coordinate within the image as a center point and randomly erase regions with a size from $1\%$ up to $25\%$ of the input size. Note, we refer to this method as patch for 2D and cube for 3D.\\ \\
Second, we extend this approach and split a single patch or cube into multiple ones. To this end, we mask-out up to ten randomly located and sized patches or cubes within an input image, while the overall erasing size remains in the limit of $1\%$ up to $25\%$ of the input size. We call this method multiple-patch or multiple-cube for 2D and 3D, respectively. \\ \\
Third, we erase entire brain sides based on the idea of stimulating the networks to exploit the symmetry of a brain. Hence, we randomly erase the right or left side of the brain in the input slice. Similar for 3D, here we randomly erase the right or left side of the brain in 1 up to 32 multiple sequential input slices. We refer to this method as half-slice for 2D and half-volume for 3D.
\\ \\
We systematically evaluate all erasing methods with different strategies for masking-out the regions. First, we simply erase regions in the input, e.g., all intensity values of a region are set to zero similar to previous works \cite{devries2017improved,pathak2016context,zimmerer2018context}. Second, to further increase the variance of our erasing methods we fill the erased region with noise sampled from the image pixel distribution.
\\
For all our methods, we set the probability of the spatial erasing to $p=0.5$, such that the network still receives unmodified images.

\subsection{Training and Evaluation}
We follow the idea of VAEs, hence we optimize our networks with respect to the reconstruction loss between the original input image and the network output reconstruction combined with the constraint that the latent variables follow a multivariate normal distribution. Hence, our loss function is based on the $l_{1}$-distance between our input and output combined with the distribution-matching Kullback–Leibler divergence for regularization. We train our networks with a batch size of 32 using Adam for optimization with a learning rate of $0.001$. We individually tune the number of training epochs of the networks using the reconstruction performance on our validation set with images of healthy subjects. \\
For all evaluations, we employ the following post-processing steps. First, we multiply each residual image by a slightly eroded brain mask to account for errors occurring at sharp brain-mask boundaries. Next, we remove small outliers with a median filter. For anomaly segmentation of a test image, we consider the voxel-wise residuals obtained from the $l_{1}$-distance between the original input image and the network's reconstruction. \\ \\
For comparison of our methods, we consider voxel-wise anomaly segmentation performance. To this end, we consider the Dice coefficient (DICE) which is defined by
\[
\mathrm{DICE}=\frac{2\left|X\cap Y\right|}{\left|X\right|+\left|Y\right|}
\]
with two sets $X$ and $Y$. Noteworthy, evaluating the DICE requires binarization of the difference image between the original input image and the network's reconstruction.
For this purpose, we utilize our validation set and perform a greedy search to determine the binarization threshold for the segmentation, similar to \cite{baur2020autoencoders}. Since the scans are normalized, intensity intervals range from 0 to 1. Using the ground truth segmentation, we compute the DICE on the validation set for thresholds at the upper and lower quartile of the center of the intensity interval. Based on the DICE we cut the interval to either the lower or upper half and continue the search with the updated interval. The procedure is repeated for 10 iterations and we use the binarization threshold that leads to the best DICE score. Afterwards, we use the determined binarization threshold for the test sets. We report the DICE on an entire data set (DICE$_{D}$), and also report mean and standard deviation for the subject-wise values (DICE$_{S}$). Moreover, to evaluate the models performance for different operating points, e.g., binarization threshold for segmentation, we also consider the area under the Precision-Recall-Curve (AUPRC). Here, for each data set, we generate Precision-Recall-Curves (PRC) for each model and then we compute the area under it (AUPRC).\\ Moreover, we consider our best performing methods and our baseline methods with respect to slice-wise anomaly detection.
This allows for localization of anomalies on a slice-level in a volume, i.e., which slice contains a lesion. For this purpose, we divided each volume in our test set into normal and abnormal slices. Considering the lesion annotations, we strictly consider all slices with annotations as abnormal and normal otherwise. For discrimination between normal and abnormal slices, we use the $l_{1}$-distance between the original input and the network's reconstruction calculated for each slice. For evaluation of our slice-wise anomaly detection performance independent of the operating point we report the AUPRC.

\section{Results}
\begin{figure}
\centering
\includegraphics[width=0.44\textwidth]{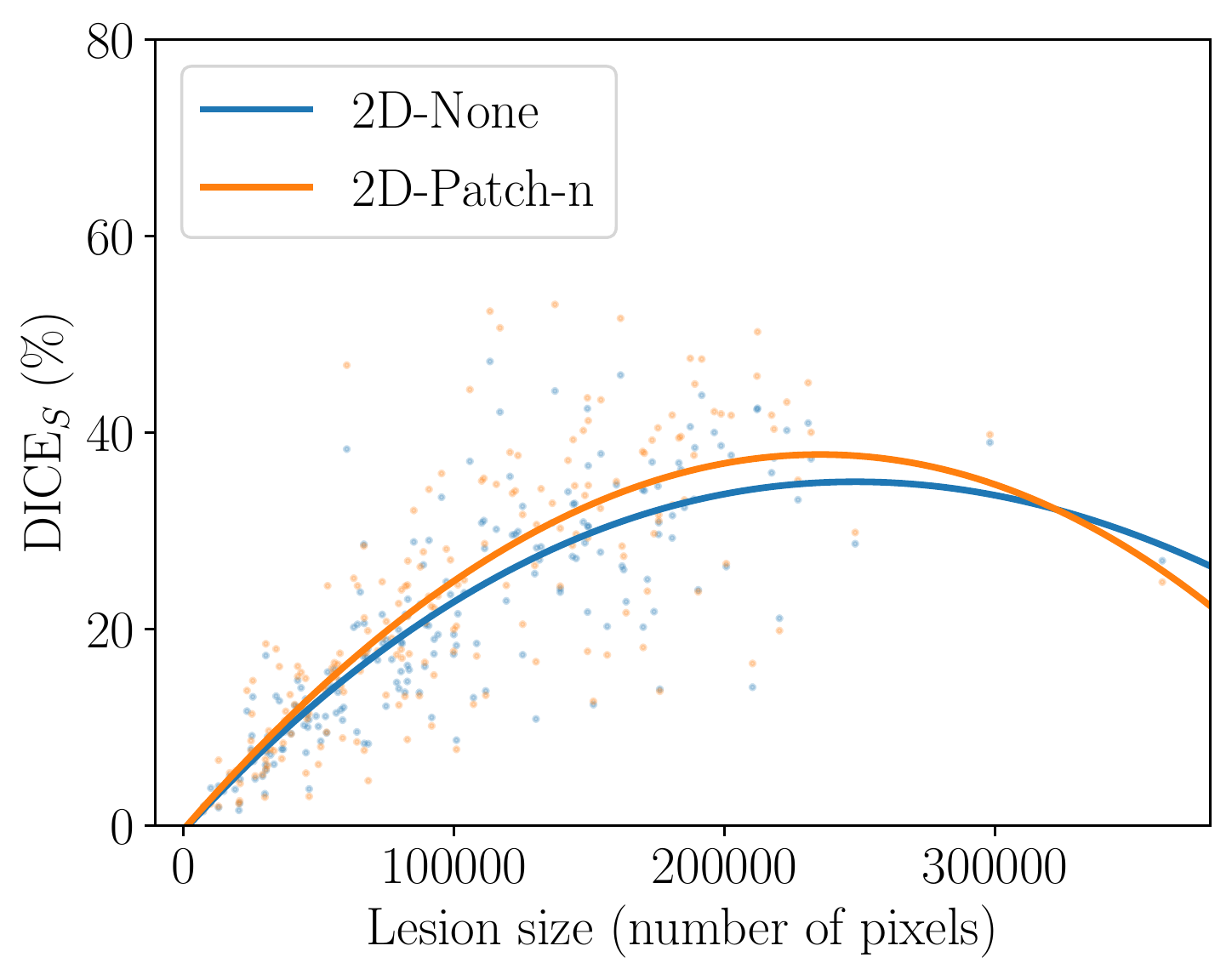}
\includegraphics[width=0.44\textwidth]{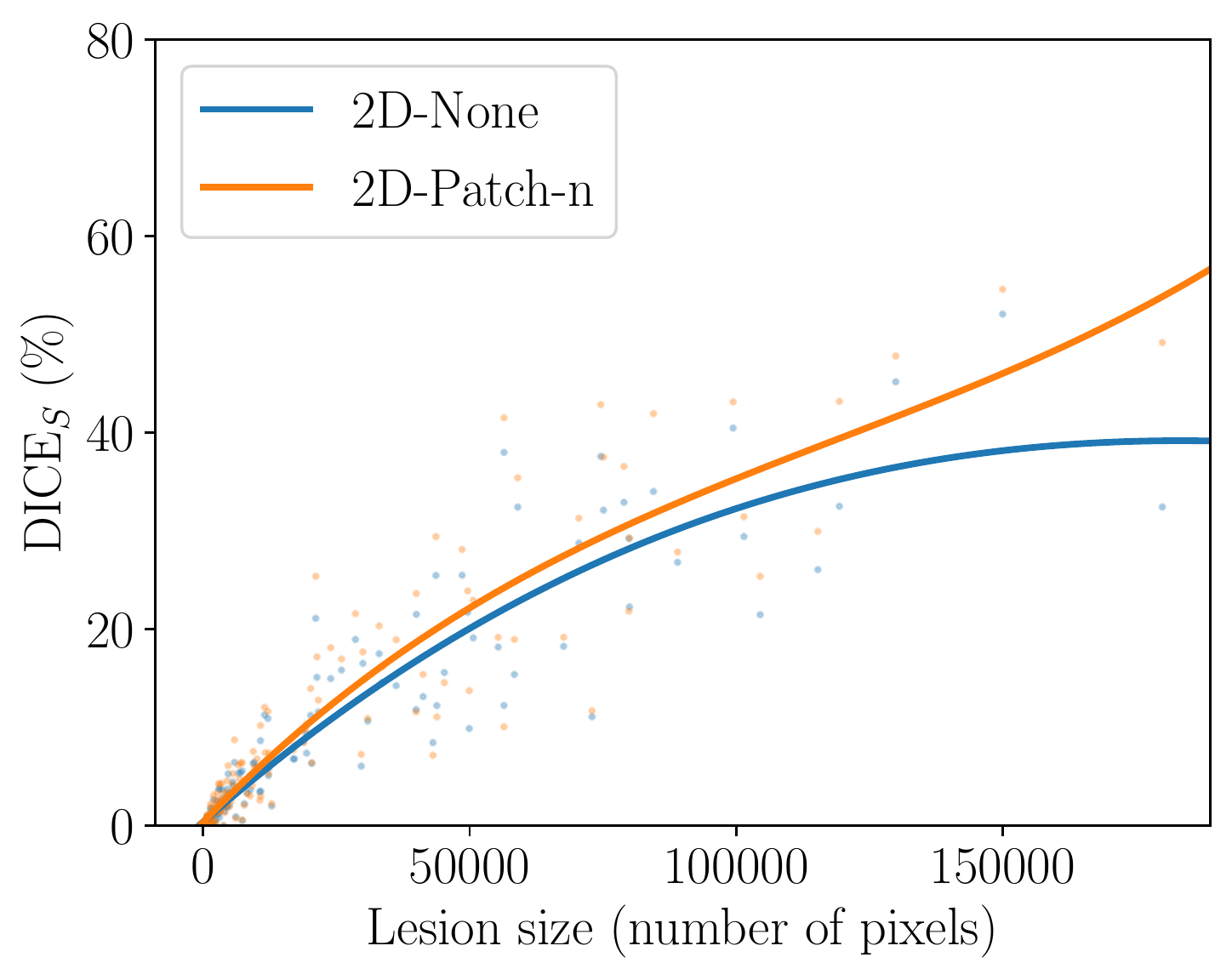} \\
\includegraphics[width=0.44\textwidth]{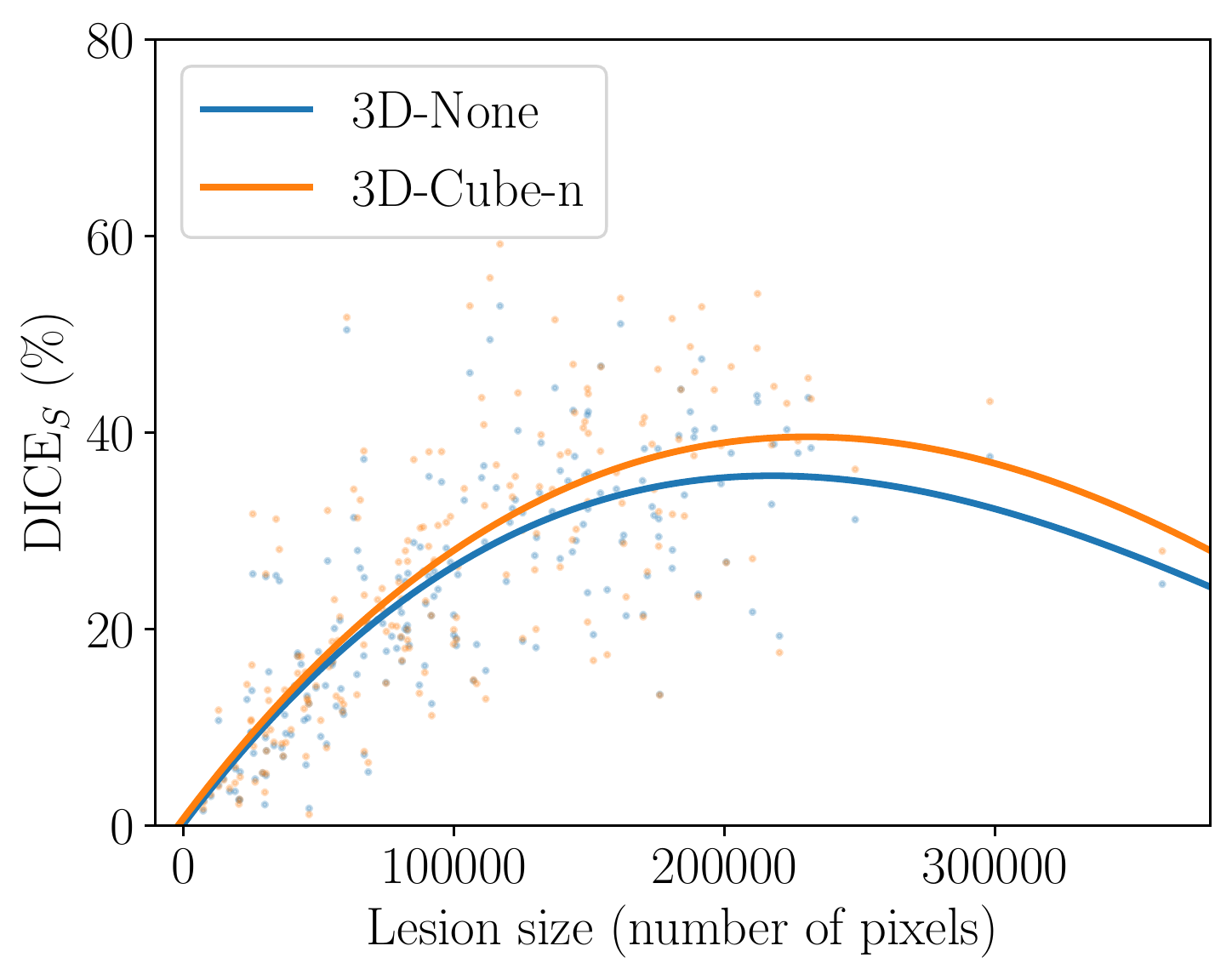} 
\includegraphics[width=0.44\textwidth]{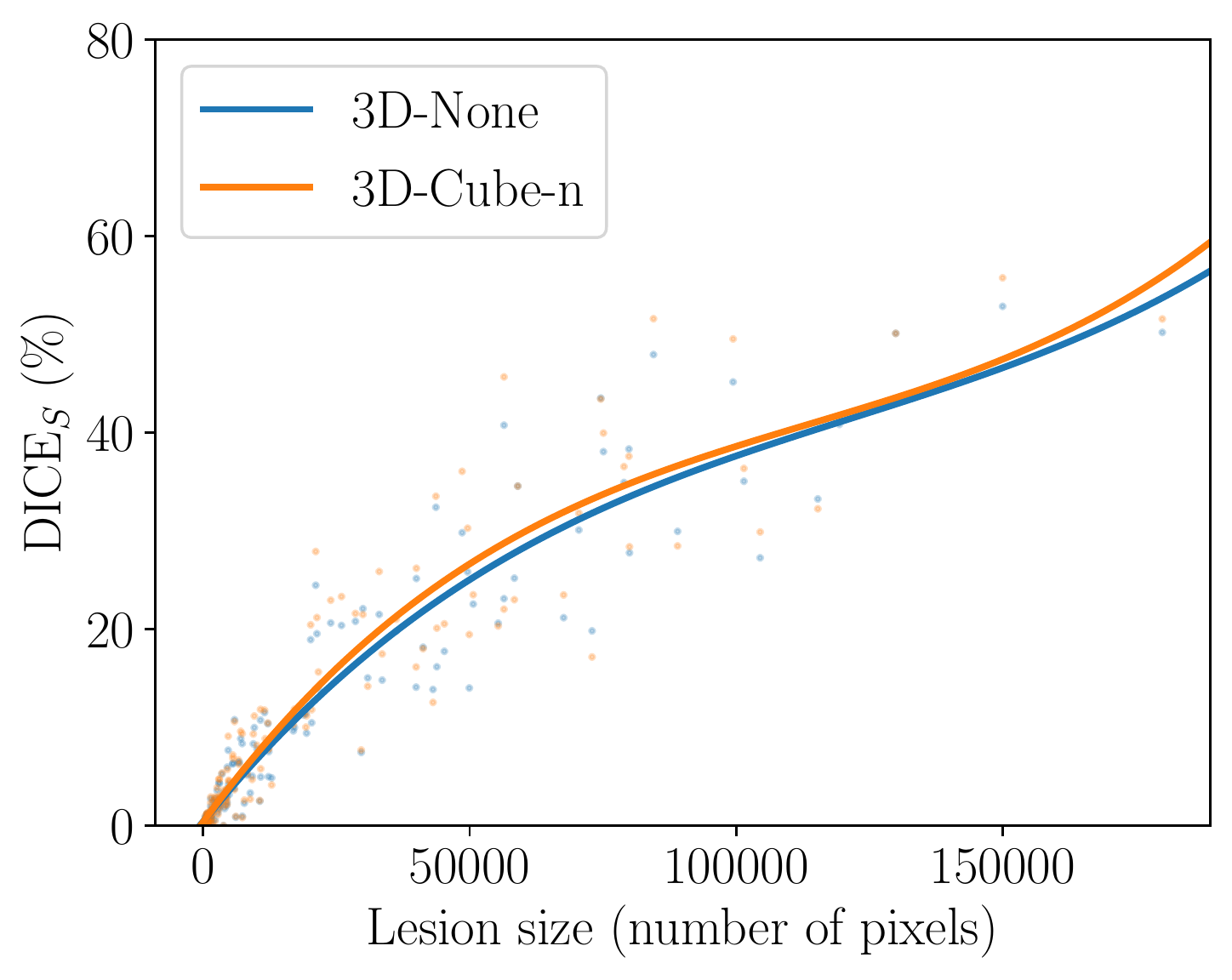} \\
\includegraphics[width=0.44\textwidth]{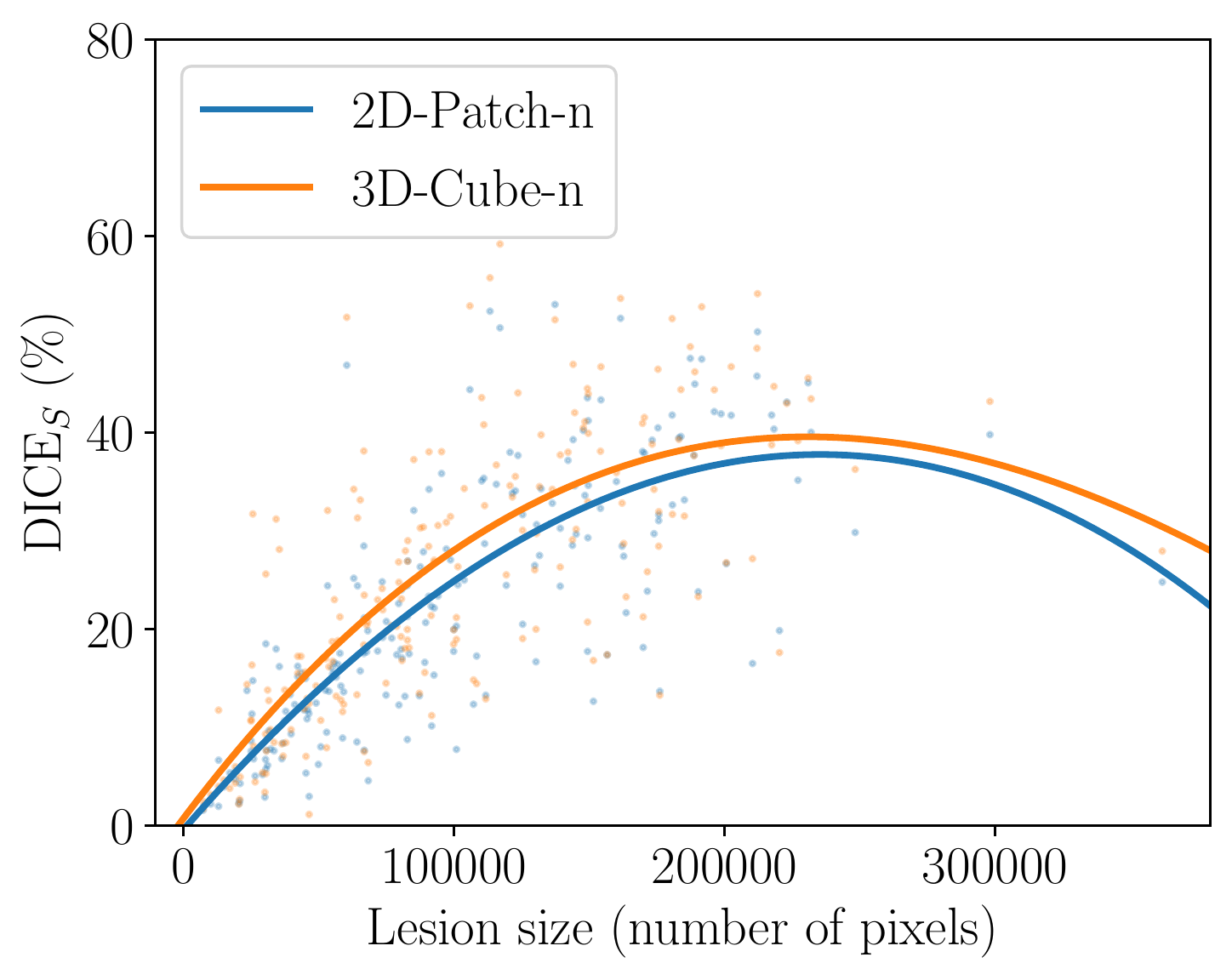}
\includegraphics[width=0.44\textwidth]{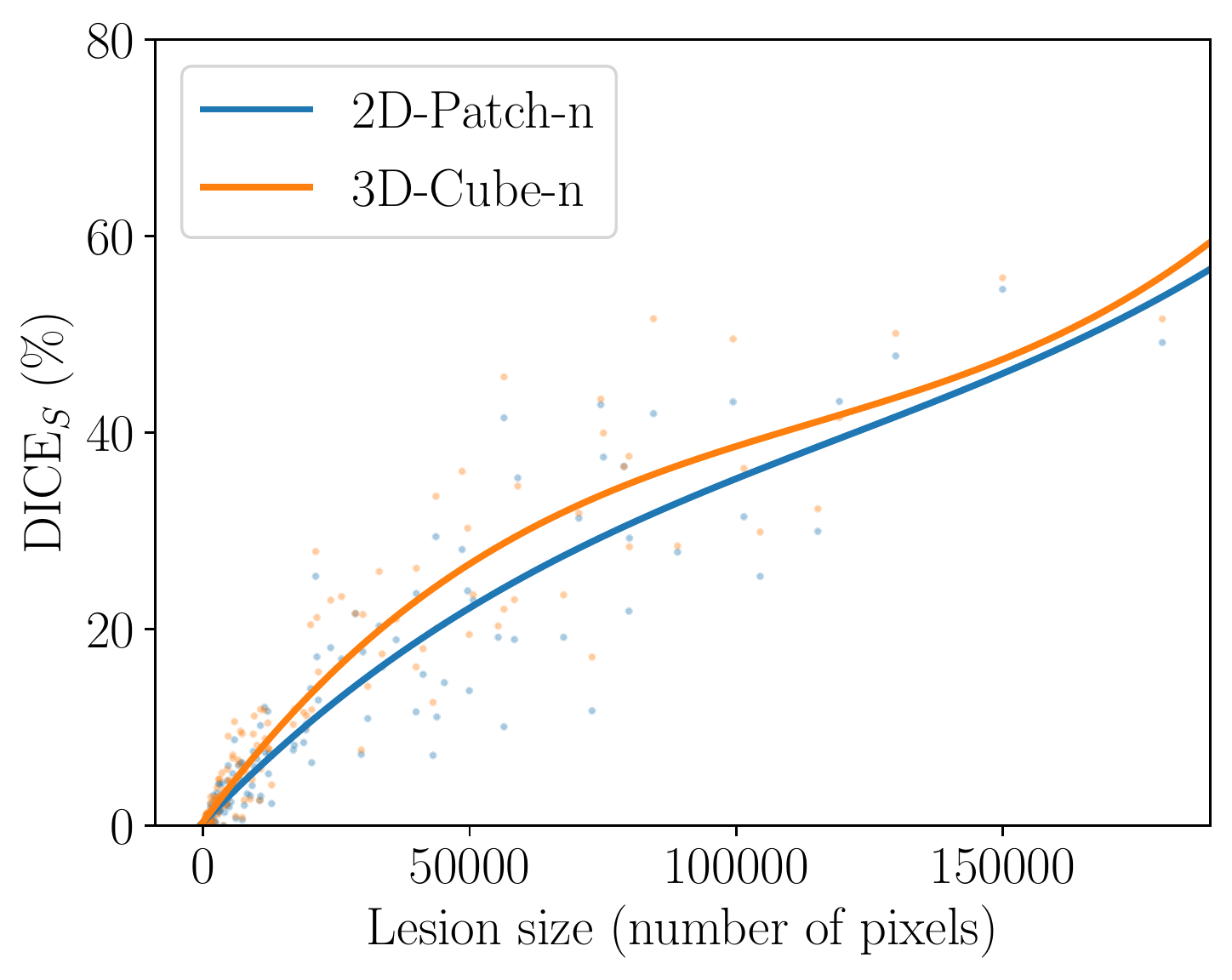}

\caption{Subject-wise DICE$_{S}$ over lesion size. Lesion size refers to the number of annotated pixels for the lesion. Results for the BraTS 2019 data set and ATLAS data set are shown left and right, respectively. (Top) Comparing 2D VAE with and without erasing; (Middle) Comparing 3D VAE with and without erasing; (Bottom) Comparing 2D and 3D VAE with erasing. Transparent dots refer to the subject-wise DICE$_{S}$ scores. Solid lines are derived by a polynomial regression of order three.}
\label{fig:DiceOLesion}
\end{figure}

\begin{figure}
\centering
\includegraphics[width=0.9\textwidth]{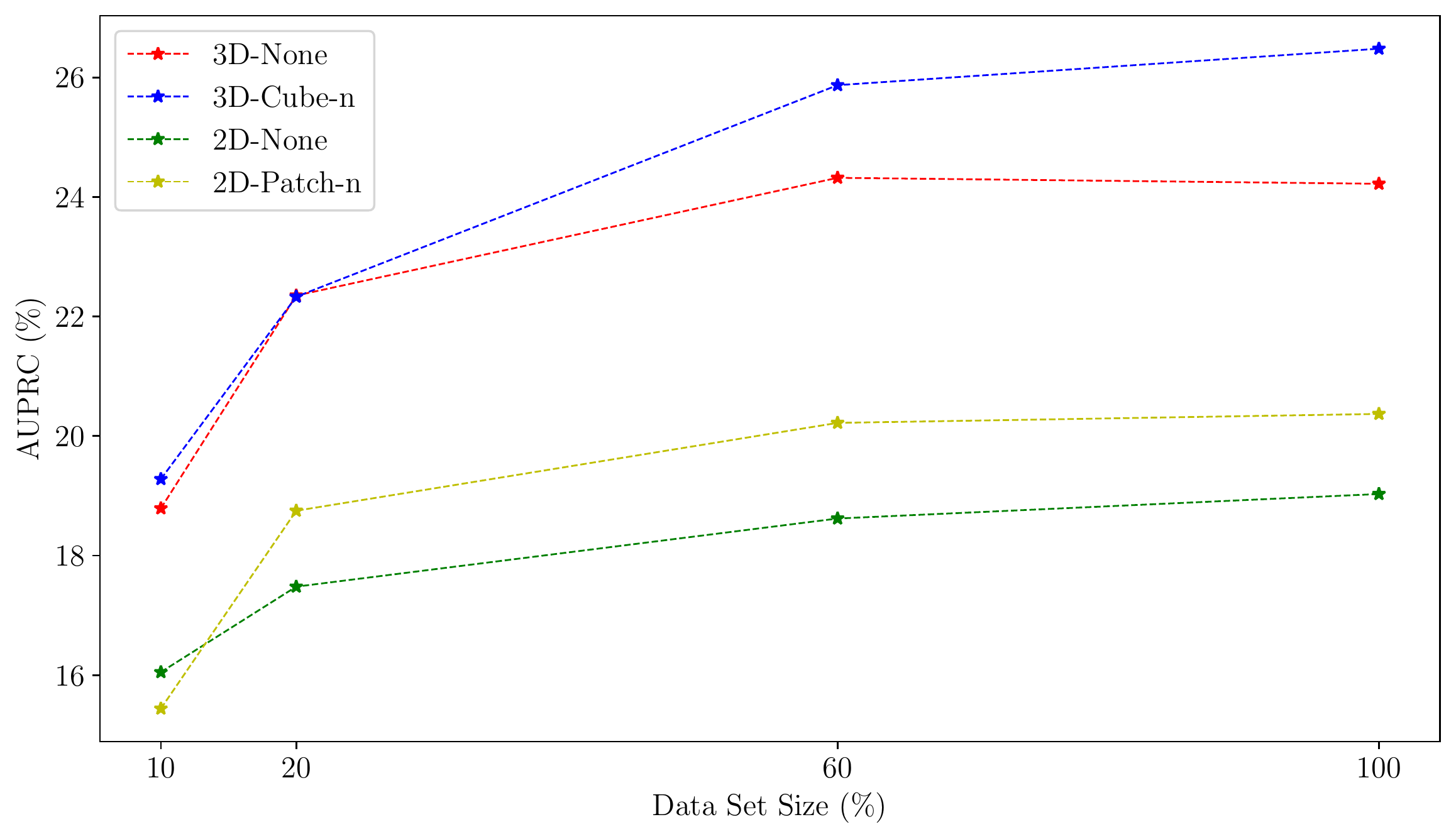}
\caption{Impact of data set size on the UAD performance. We train our methods with 10\%, 20\%, 60\% and 100\% of the training data, shown is the average AUPRC using our two test data sets (BraTS 2019, ATLAS).}
\label{fig:results_data_set_size}
\end{figure}

\begin{figure}
\centering
\includegraphics[width=0.58\textwidth]{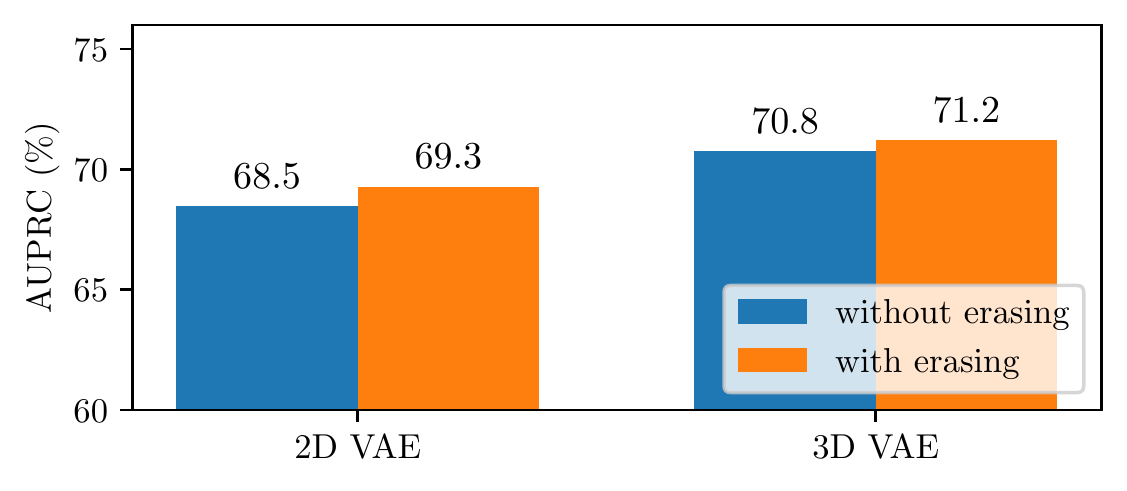}
\caption{Slice-wise anomaly detection for our baseline and best performing methods. Shown is the AUPRC on the combination of our test sets (BraTS 2019, ATLAS). 2D VAE with and without erasing refers to 2D-None and 2D-Patch-n, respectively. 3D VAE and without erasing refers to 3D-None and 3D-cube-n, respectively.}
\label{fig:slice_wise_detection}
\end{figure}

\begin{figure}
\centering
\includegraphics[width=1.0\textwidth]{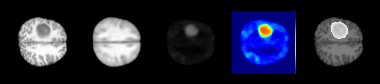}
\includegraphics[width=1.0\textwidth]{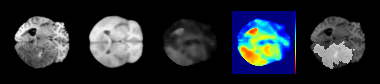}
\vspace{0.5cm} \\
\includegraphics[width=1.0\textwidth]{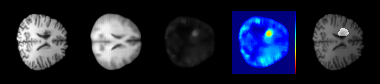}
\includegraphics[width=1.0\textwidth]{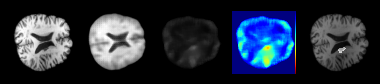}

\caption{Four example test cases using our best performing method 3D-cube-n. From left to right: Input image, output image, difference image, heat-map difference image, and ground truth segmentation. The first two lines contain examples from the BraTS 2019 data set and the two the two bottom lines contain examples from the ATLAS data set.}
\label{fig:results_example_images}
\end{figure}

First, we compare 2D and 3D UAD deep learning methods combined with our erasing regularization methods in Table \ref{tab:All-networks-with metrics}. For both VAEs, our different erasing methods lead to performance improvements. Overall, our 3D VAE outperforms the 2D VAE for all our experiments. Using noise for masking-out the regions works slightly better than masking-out with zeros. For our 3D VAE using a single cube for erasing, followed by masking-out an entire brain side in a subvolume works best. Considering our 2D-VAE, masking-out an entire brain side shows the best results, closely followed by masking-out multiple patches.  Comparing the DICE$_{D}$ of our best performing 3D approach (3D-cube-n) with the 2D baseline approach (2D-None) demonstrates a relative performance improvement of $12.31\%$ and $32.20\%$ on the BraTS 2019 and ATLAS data set, respectively. \\ Second, we evaluate the performance of our baselines and best performing methods with respect to lesion size in Figure \ref{fig:DiceOLesion}. Here, our results demonstrate that the smallest and largest lesions are challenging. Consistently, using erasing improves the DICE$_{S}$ over all lesion sizes, while being particularly effective for large lesions. Also, comparing 2D and 3D methods shows that 3D consistently outperforms 2D, especially for small lesions. \\ Third, we evaluate the effect of the data set size in Figure \ref{fig:results_data_set_size}. Reducing the data set has a pronounced impact on the performance for 3D as well as 2D, especially when less than 60\% of the training data is used. Also, the spatial erasing works better when the network is trained with more data. While reducing the data set size has a larger impact on 3D, even with only 20\% of the training data the 3D VAE works better than the 2D VAE with erasing and 100\% of the training data. Moreover, our erasing turns out to be effective for the 2D VAE, considering that a 2D VAE without erasing trained with 100\% of data is outperformed by a 2D VAE with erasing trained with only 20\% of the data. \\Fourth, Figure \ref{fig:results_example_images} demonstrates example images for our best performing method 3D-Cube-n. Notably, the ground truth segmentation are highlighted in all difference images, while also showing errors at further regions. \\ Moreover, we use our best performing 2D and 3D methods trained on T1-weighted MRI data and evaluate on T1ce-weighed MRI data from the BraTS 2019 data set to study the effect of using additional image information, see Table \ref{tab:t1vst1ce}. Here, we observe immediate performance improvements compared to T1-weighting for both 2D and 3D with a relative improvement of 13.61\% and 21.82\% for 2D and 3D considering the DICE$_{D}$.\\ Last, we evaluate our baseline and best performing methods with respect to slice-wise anomaly detection, see Figure \ref{fig:slice_wise_detection}. Here, our best performing method achieves an AUPRC of 71.2\%. Also for this task using 3D information and erasing turns out to be beneficial, improving the AUPRC by approximately 4\% compared to the 2D VAE.

\label{Results}

\begin{table} 
\centering
\caption{Results for our 2D and 3D VAE combined with our spatial erasing methods evaluated on the BraTS 2019 and ATLAS (Stroke) data set. The abbreviations for input and erasing refer to the input/VAE dimension, erasing strategy and value used for masking-out a region, e.g., 2D-Patch-0 and 2D-Patch-n stand for a 2D VAE with patch erasing, while the first refers to masking-out a region with zeros and the second refers to masking-out a region with noise. DICE$_{D}$ represents the metric based on the voxel calculation of an entire data set. DICE$_{S}$ ($\mu\pm \sigma$) refers to the mean and standard deviation of the subject-wise score. All metrics are in percent.} 
\label{tab:All-networks-with metrics}%
\begin{tabular}{lcccc}
    \multicolumn{4}{c}{BraTS  2019}   \\ 
     \hline
  Input \&  Erasing  & DICE$_{D}$ & DICE$_{S}$ ($\mu\pm \sigma$)\  & AUPRC \\
  \hline
  2D-None  & 26.80&25.30 $\pm$ 12.37 & 21.19  \\
  3D-None & 28.14&26.93 $\pm$ 12.40 & 24.69  \\
 \hline
  2D-Patch-0 & 27.96& 26.52 $\pm$ 13.42 & 22.53   \\
  2D-Patch-n  & 27.99&26.58 $\pm$ 13.27 &	22.54  \\
  3D-Cube-0 & 29.24& 27.90 $\pm$ 13.57 & 26.18  \\
  3D-Cube-n & 30.10& 28.80 $\pm$ 13.74 &	27.85  \\
 \hline
  2D-Multi-Patch-0 & 28.10& 26.44 $\pm$ 12.89  & 22.54  \\
  2D-Multi-Patch-n &  28.51& 27.24 $\pm$ 13.14 &	22.81 \\
  3D-Multi-Cube-0 &  28.88& 27.67 $\pm$ 13.22& 25.82 \\
  3D-Multi-Cube-n & 29.52&28.33 $\pm$ 13.42 &	26.18\\
   \hline
  2D-Half-Slice-0 &   26.86&25.44 $\pm$ 12.42 & 21.77  \\
  2D-Half-Slice-n &  27.97& 26.45 $\pm$ 13.22 &	22.84 \\
  3D-Half-Volume-0 & 28.49& 27.51 $\pm$ 13.17  & 25.47  \\
  3D-Half-Volume-n  &28.99&27.92 $\pm$ 13.24 &	26.07  \\
 \hline
\hline
  \end{tabular}

\centering
\vspace{5mm}
\begin{tabular}{lcccc}

    \multicolumn{4}{c}{ATLAS (Stroke)}   \\ 
     \hline
  Input \&  Erasing  &  DICE$_{D}$ & DICE$_{S}$ ($\mu\pm \sigma$)\  & AUPRC \\
  \hline
  2D-None  & 24.72& 11.23 $\pm$ 13.66 & 16.86  \\
  3D-None & 30.68& 14.42 $\pm$ 16.06 & 23.74  \\ 
  \hline
  2D-Patch-0 & 27.68& 12.23 $\pm$ 13.67 & 18.65 \\
  2D-Patch-n  & 27.42&12.36 $\pm$ 14.61 & 18.20 \\
  3D-Cube-0 & 31.50 & 15.59 $\pm$ 17.02 & 23.47 \\ 
  3D-Cube-n& 32.68 &15.53 $\pm$ 17.30 & 25.11  \\ 
 \hline
  2D-Multi-Patch-0 &26.99&11.82 $\pm$ 14.29 &18.72  \\
  2D-Multi-Patch-n& 28.06& 12.88 $\pm$ 15.21 & 19.49 \\
  3D-Multi-Cube-0 & 31.83 & 15.23 $\pm$ 16.64 & 24.51 \\ 
  3D-Multi-Cube-n  & 32.37& 14.99 $\pm$ 17.31 & 25.13 \\ 
   \hline
  2D-Half-Slice-0 & 27.54& 11.05 $\pm$ 13.70 & 18.60 \\
  2D-Half-Slice-n & 28.99& 12.13 $\pm$ 14.79  & 20.37  \\
  3D-Half-Volume-0 & 31.00& 15.21 $\pm$ 17.00 & 23.14 \\ 
  3D-Half-Volume-n & 33.05 & 15.27 $\pm$ 17.21  & 25.58 \\ 
 \hline
\hline
  \end{tabular}
\end{table}

\begin{table} 
\centering
\caption{Results for additional image information considering the BraTS 2019 data set. DICE$_{D}$ represents the metric based on the voxel calculation of an entire data set. DICE$_{S}$ ($\mu\pm \sigma$) refers to the mean and standard deviation of the subject-wise score. All metrics are in percent.} 

\label{tab:t1vst1ce}%

\begin{tabular}{lccccc}

  Input \&  Erasing  & Sequence & DICE$_{D}$& DICE$_{S}$ ($\mu\pm \sigma$)  & AUPRC &   \\ 
  \hline
  2D-Patch-n  & T1 & 27.99 &26.58 $\pm$ 13.27&	22.54    \\
  2D-Patch-n  & T1ce & 31.80 & 29.08 $\pm$ 12.77 & 24.28   \\
    \hline
  3D-Cube-n & T1 & 30.10 & 28.80 $\pm$ 13.74 &	27.85   \\ 
  3D-Cube-n & T1ce & 36.67& 33.40 $\pm$ 14.55  & 31.12   \\ 
        
 \hline
\hline
  \end{tabular}
\end{table}

\section{Discussion }
\label{Discussion}
We consider the problem of unsupervised anomaly segmentation and propose to learn from entire 3D MRI volumes instead of single 2D MRI. For this purpose, we extend 2D VAEs to 3D and also propose several different input erasing methods for regularization. 
Comparing our 2D VAE (2D-None) with the corresponding 3D version (3D-None) without any input erasing demonstrates that 3D outperforms the 2D version on two public data sets, especially for the stroke data set with a DICE$_{D}$ of $30.86\%$ for 3D compared to a DICE$_{D}$ of $24.72\%$ for 2D, see Table \ref{tab:All-networks-with metrics}. This highlights that 3D information can be effectively leveraged by a 3D VAE and agrees with our expectation that increased spatial context by using entire MRI volumes allows for improved anomaly segmentation performance. \\ \\
We also evaluate 2D and 3D input erasing for regularization and train the networks to restore missing image parts conditioned on its surroundings. Our results in Table \ref{tab:All-networks-with metrics} demonstrate that input erasing allows for further performance improvements both for our 2D and 3D VAE. Regarding the method for masking-out a region, previous works in 2D mostly simply mask our input regions with zeros \cite{zimmerer2018context,pathak2016context,devries2017improved}. However, our results demonstrate that using noise for masking-out a region in the input works slightly better, indicating that the increased variance during training is advantageous for regularization. \\
We also consider different strategies such as erasing multiple patches or an entire brain side. While all erasing strategies are beneficial, there is no clear winner between the different strategies considering our results on both data sets. 
Furthermore, one could argue that our input erasing leads to brain anatomy that deviates from normal, which is in slight contrast to the idea of only providing healthy brain anatomy as input. However, our ground-truth image that is used for optimization remains unmodified, hence our networks are enforced to solve an in-painting task for abnormal regions. Our results demonstrate that this leads to an improved segmentation performance. \\ \\ To gain further insights, we study the performance with respect to the lesion size in Figure \ref{fig:DiceOLesion}. While providing consistent performance improvements, erasing turns out to be especially valuable for larger lesions. This might be attributed to the fact that with erasing, networks are enforced to solve an additional in-painting task, making them suited to handle inputs with large anomalies. Also, our results in Figure \ref{fig:DiceOLesion} further emphasize the value of 3D information, especially for smaller lesions considering the ATLAS data set.\\ \\ Next, we study the effect of the training data set size. As expected, the data set size has a notable impact on the performance, see Figure \ref{fig:results_data_set_size}. It stands out that our 3D methods trained with only 20\% of the training data even outperform the 2D methods trained with 100\% of the data. This indicates that increasing the spatial context during training is even more important than increasing the data set size. This is an interesting observation, as one could assume that due to the increased number of parameters, 3D-Models require more data compared to their 2D-counterparts. We believe that this counter-intuitive behaviour could explained by the increased complexity of the task and the bigger input image for the 3D approach. The learning task of the 3D model can be considered more complex since an entire volume must be processed and reconstructed at once, while 2D is only trained to process a single slice. Also, for 3D the input image is bigger (volume) compared to 2D (single slice). Note, if the input image is bigger, then a network might need more expressive power to capture the patterns in the input image, as shown in \cite{tan2019efficientnet}. \\ Considering our erasing approach and the data set size suggests that solving the additional in-painting task needs sufficient training data to provide effective regularization. However, with only 60\% of the training data our models with our regularization approach lead to higher performance than a model without regularization trained with the full dataset. We argue this demonstrates the effectiveness of our regularization approach, as less data is required to achieve similar or better performance compared to a model without regularization. Still, increasing the data set size is valuable as the performance for our model with erasing continues to improve with a larger training data set. 
\\ \\
Comparing our novel 3D methods with input erasing with the previous 2D approach demonstrates a relative performance improvement of $12.31\%$ and $32.20\%$ on the BraTS 2019 and ATLAS data set, respectively. A comparable work evaluating UAD performance on the same ATLAS data set achieves a mean subject-wise DICE score of $12 \pm 12 \%$ with their best performing method \cite{chen2020unsupervised}. Notably, this 2D method is restoration-based and involves significantly increased computational complexity. Our 3D approach with input erasing leads to a mean subject-wise DICE score of $15.53\pm17.30\%$, improving the UAD state-of-the-art on this data set. This demonstrates the effectiveness of our approach. Comparing our results on the BraTS 2019 data set with other works that utilize additional image information, e.g. T2-weighted data \cite{chen2020unsupervised,zimmerer2018context}, highlights the advantage of additional image information. Similar, we observe immediate performance improvement for our methods when evaluated on T1ce-weighted data, despite the domain adaption from T1, see Table \ref{tab:t1vst1ce}. Also, other studies that use multiple MRI sequences \cite{baur2020autoencoders,baur2018deep} achieve higher performance metrics, however, a direct comparison is difficult due to different data sets and settings. Notably, multiple MRI sequences are beneficial but not always available \cite{ito2019comparison,akkus2017deep}, imposing an additional challenge on UAD. \\ Putting UAD into perspective with supervised methods demonstrates that segmentation performance is in a moderate range. Considering the BRATS 2019 data set, supervised methods achieve a mean subject-wise DICE score of around 90\% \cite{jiang2019two} utilizing all available MRI sequences (T1, T1ce, T2, FLAIR). Considering the ATLAS data set, supervised methods achieve mean subject-wise DICE scores in the range of 32.92\% up to 53.49\% \cite{ito2019comparison}. While UAD is notably more challenging than supervised segmentation, the overall UAD performance on these supervised data sets might also be limited, as the annotation focuses on pre-specified lesions and not all anomalies in the images might be labeled. This is also demonstrated in Figure \ref{fig:results_example_images}, where, e.g., the segmentation focuses only on the tumor and not on all brain regions that deviate from normal. Also, the domain shifts between different data sets might be challenging, which is also pointed out in previous works \cite{baur2020autoencoders,zimmerer2018context}.\\ Considering these challenges we also evaluate our methods with respect to slice-wise anomaly detection, see Figure \ref{fig:slice_wise_detection}. Here, we observe significantly increased performance compared to segmentation with an AUPRC of 71.2\% for our best performing method.  The slice-wise detection performance motivates that UAD can be helpful in red-flagging suspicious MRI data in clinical routine, especially with T1-weighted MRI data. Also, we believe that unsupervised segmentation gives additional cues to the reader as to where an anomaly may be located and thus it is helpful to quickly localize a potential anomaly or lesion. For this our work consist a valuable contribution by demonstrating the benefits and emphasizing the use of 3D-models with spatial erasing for voxel-wise and slice-wise UAD.  \\ \\ 
For future work, our findings could be extended to more complex deep learning methods for UAD, such as GANs \cite{schlegl2019f}. In particular, combining our 3D approach with restoration-based methods \cite{chen2020unsupervised} might improve the overall performance. However, this approach also leads to significantly increased runtime and computational efforts, e.g., a restoration accumulates quickly to multiple minutes for a single MRI \cite{baur2020autoencoders}, which is particularly challenging for clinical routine.

\section{Conclusion }
\label{Conclusion}
We study the task of unsupervised anomaly segmentation in brain MRI and propose to use entire 3D MRI volumes instead of single 2D MRI slices by extending 2D VAEs to 3D.  Also, we study and extend the concept of input erasing and propose several different 3D input erasing strategies for regularization. Overall, our results demonstrate that using increased spatial context by using entire MRI volumes combined with 3D deep learning clearly outperforms 2D methods. Also, we observe that combining deep learning with spatial input erasing allows for further performance improvements and reduces the requirement for large training data sets.

\section*{Compliance with ethical standards}
\begin{small}
\textbf{Funding:} This work was partially funded by Grant Number ZF4026303TS9 \\ 
\textbf{Conflict of interest:} The authors declare that they have no conflict of interest.\\ 
\textbf{Ethical approval:} This work was conducted retrospectively on data from clinical routine which was completely anonymized. Ethical approval was therefore not required. Also this work relies on the BraTS 2019 and ATLAS data set. For use of these data sets, no ethics statements are necessary. \\ 
\textbf{Informed consent:} Not applicable
\end{small}

\bibliographystyle{spmpsci}      
\bibliography{CARS_Tracking.bib}   

\end{document}